\documentclass[a4paper]{article}%
\usepackage{amsmath}
\usepackage{amsfonts}
\usepackage{amssymb}
\usepackage{graphicx}%
\providecommand{\U}[1]{\protect\rule{.1in}{.1in}}
\begin{document}

\title{Null-Result\textrm{\ }Detection\ and Einstein-Podolsky-Rosen Correlations}
\author{Luiz Carlos Ryff\\\textit{Instituto de F\'{\i}sica, Universidade Federal do Rio de Janeiro, }\\\textit{Caixa Postal 68528, 21041-972 RJ, Brazil}\\E-mail{\small : ryff@if.ufrj.br}}
\maketitle

\begin{abstract}
It follows from Bell's theorem and quantum mechanics that the detection of a
particle of an entangled pair can (somehow) \textquotedblleft
force\textquotedblright\ the other distant particle of the pair into a
well-defined state (which is equivalent to a reduction of the state vector):
no property previously shared by the particles can explain the predicted
quantum correlations. This result has been corroborated by experiment,
although some loopholes still remain. However, it has not been experimentally
proved -- and it is far from obvious -- that the absence of detection, as in
null-result (NR) experiments could have the very same effect. In this paper a
way to try to bridge this gap is suggested. As already shown for the case of
Einstein-Podolsky-Rosen (EPR)\textrm{ }correlations, if NR detections cannot
induce a reduction of the state vector, then faster-than-light (FTL)
communication becomes possible, at least in principle. But, it will be
demonstrated that -- as entertained by Bohm -- this does not necessarily lead
to a causal paradox, or to the rejection of the Lorentz transformations.

\end{abstract}

\bigskip\textbf{I. Introduction}

\bigskip

In a\textrm{ }null-result (NR)\ experiment, instead of registering the
presence of a particle, a\textrm{ }detector \textquotedblleft
registers\textquotedblright\ its absence \textrm{[1]}. In some situations,
since there is no detector \textquotedblleft click,\textquotedblright\ it is
possible to know the path that the particle has not followed and, by
exclusion,\textrm{ }we can also infer the path followed by the particle.
Naturally, from the habitual standpoint there seems to be nothing mysterious
in this. On the other hand, although they involve no irreversible
amplification (that is usually associated with the measurement process
\textrm{[2]}), it is accepted that NR detections have the same capability of
reducing (or collapsing) the quantum state vector as ordinary (O)\ detections.
However, as has already been argued, the experiments so far discussed can be
explained without the need to invoke any collapse \textrm{[3]}. It is also
noteworthy that although the experimental violations of Bell's inequalities
\textrm{[4]} corroborate the point of view according to which the detection of
a particle of an entangled pair \textquotedblleft forces\textquotedblright%
\ the other distant particle of the pair into a well-defined state (more
specifically, no property previously shared by the particles can explain the
observed quantum correlations), it has not been proved--and it is far from
obvious--that the absence of detection, as in NR measurements, could have the
very same effect. Therefore, the subject, although relatively old, is far from
being settled. A way of trying to bridge this gap will be suggested here. This
approach differs from previous ones \textrm{[5,6]} in two aspects: (1)
time-like events will be considered (and the importance of doing so will be
stressed), and (2) coincident detections will be taken into account, which
simplifies the experiment and makes the differences between the predictions
more accentuated, based, on the one hand, on the assumption that NR detection
induces a collapse and, on the other, on the assumption that NR detection does
not induce a collapse\textrm{ }(at least as it is usually understood). As
already shown \textrm{[5,6]}, if NR detections cannot induce, via
Einstein-Podolsky-Rosen\textrm{ }(EPR) correlations \textrm{[7]}, a reduction
of the state vector, then, faster-than-light (FTL) communication (an idea
entertained by Bohm \textrm{[8]}) becomes possible, at least in principle. In
this paper it will be demonstrated that, strange as it may sound, this can be
consistent with Lorentz transformations, not leading to any causal paradox.

To see the importance of considering time-like events, a point that has been
overlooked in previous discussions on NR detection, let us consider the
following simple experiment. A single photon (from a pair generated via
spontaneous parametric down-conversion (SPDC) \textrm{[9]}, for instance),
impinges on a 50:50 beam splitter. A first detector is placed near the beam
splitter to register a reflected photon, and a second one is placed distant
from the beam splitter to register a transmitted photon. Whenever the first
detector does not click, it is possible to infer that the photon has been
transmitted. From a quantum mechanical standpoint, it can be said that the
absence of detection induced a reduction of the state vector,
\textquotedblleft forcing\textquotedblright\ the photon into the transmitted
state. However, the lack of detection at the first detector and the detection
at the second are space-like events \textrm{[10]}; therefore, there are an
infinite number of Lorentz frames in which the second detector clicks before
the first has not clicked. Hence, it can equally be said that the detection of
the second photon induced the collapse of the reflected state. Consequently,
this cannot be considered an indisputable NR-detection measurement.

Actually, the above example also admits a simple interpretation: in an ideal
situation, if a particle has not been reflected, it has necessarily been
transmitted \textrm{[11]}. To avoid this sort of explanation, it is important
to consider an experiment that violates a Bell inequality, namely one in which
an objective (as opposed to subjective) change of probability must necessarily
occur \textrm{[12]}.

As has been pointed out \textrm{[3, 6]}, in\textrm{ }the Gedanken experiments
discussed by Renninger \textrm{[13]} and Dicke \textrm{[14] }the lack of
detection, despite giving us information about the state of the system, does
not necessarily imply that the system has been forced into this state by this
very lack of detection. The same is true, for example, when we infer, from the
absence of resonance fluorescence, that a quantum jump has occurred
\textrm{[3]. }Moreover, it is always possible to describe these experiments in
another Lorentz frame in which the reduction of the state vector has not been
induced by the absence of detection.\textrm{ }Therefore, it seems fair to say
that no unquestionable collapse-inducing NR-detection measurement has been
performed since the emergence of quantum mechanics (QM), more than eighty
years ago. Naturally, NR-detection reduction of the state vector is far from
being a trivial fact, and its experimental verification is extremely important
for the investigation of the foundations of QM\textrm{ [15]}.

\newpage

\bigskip\textbf{II. The experiment}

\bigskip

Let us consider the experiment represented in Fig.1, which is a variant of an
experiment performed by Aspect, Grangier, and Roger \textrm{[16]}. To have a
clear understanding of this proposal, we will initially examine the ideal
situation. A source $S$ generates pairs of photons ($\nu_{1},\nu_{2}$) in the
state%
\begin{equation}
\mid\psi\rangle=\frac{1}{\sqrt{2}}\left(  \mid a\rangle\mid a\rangle+\mid
a_{\perp}\rangle\mid a_{\perp}\rangle\right)  , \tag{1}%
\end{equation}%
\[%
%TCIMACRO{\FRAME{itbpFU}{4.7876in}{2.1395in}{0in}{\Qcb{{}}}{}{diagramax.eps}%
%width 4.7876in;  height 2.1395in;  depth 0in;  original-width 4.734in;
%original-height 2.0998in;  cropleft "0";  croptop "1";  cropright "1";
%cropbottom "0";  filename 'diagramax.eps';file-properties "XNPEU";}} }%
%BeginExpansion
{\parbox[b]{4.7876in}{\begin{center}
\includegraphics[
height=2.1395in,
width=4.7876in
]%
{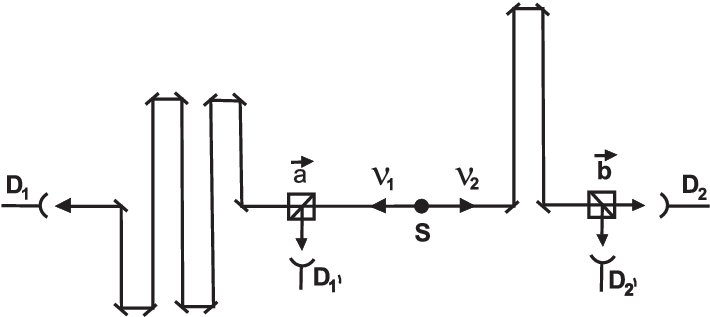}%
\\
{}%
\end{center}}}
%EndExpansion
\]
Fig.1: {\small The proposed experiment: whenever }$\nu_{1}$ {\small is not
detected at }$D_{1^{\prime}}${\small , but has not yet been detected at
}$D_{1}$, $\nu_{2}$ {\small may be forced into a well-defined polarization
state, or not, depending on the capability of NR detections to induce a
reduction of the state vector.} \smallskip\medskip\newline\smallskip where
$\mid a\rangle$ ($\mid a_{\perp}\rangle$) represents a linear polarization
state parallel (perpendicular) to $\mathbf{a}$. $\nu_{1}$ and $\nu_{2}$ are
sent in opposite directions: $\nu_{1}$ impinges on a two-channel polarizer
oriented parallel to $\mathbf{a}$, and $\nu_{2}$ (after following a detour) on
a two-channel polarizer oriented parallel to $\mathbf{b}$. Photons $\nu_{1}$
that are transmitted have to follow a detour, before impinging on detector
$D_{1}$. Photons $\nu_{1}$ that are reflected impinge on detector
$D_{1^{\prime}}$. Photons $\nu_{2}$ that are transmitted (reflected) impinge
on detector $D_{2}$ ($D_{2^{\prime}}$). The distances are such that,
independently of the Lorentz frame we use to describe the experiment, $\nu
_{1}$ is always detected (or not detected) at detector $D_{1^{\prime}}$ before
$\nu_{2}$ is detected at $D_{2}$ or $D_{2^{\prime}}$, and $\nu_{2}$ is always
detected at $D_{2}$ or $D_{2^{\prime}}$ before $\nu_{1}$ is detected at
$D_{1}$.\textsl{\ }Therefore, the detection of\textrm{\ }$\nu_{1}%
$\ at\textrm{\ }$D_{1}$\textrm{\ }cannot force\textrm{\ }$\nu_{2}$%
\textrm{\ }into a well-defined polarization state; similarly, the detection of
$\nu_{2}$\textrm{\ }at\textrm{\ }$D_{2}$\textrm{\ }or\textrm{\ }$D_{2^{\prime
}}$\textrm{\ }cannot force\textrm{\ }$\nu_{1}$\textrm{\ }into a well-defined
polarization state, since\textrm{\ }$\nu_{1}$ has already either been or not
been detected at\textrm{\ }$D_{1^{\prime}}$; the only possibility is the
detection (or non-detection) of\textrm{\ }$\nu_{1}$ at $D_{1^{\prime}}$
forcing\textrm{\ }$\nu_{2}$\textrm{. }

If, as generally believed, NR detections are capable of inducing the reduction
of the state vector, we must have%
\begin{equation}
p\left(  a_{\perp},b\right)  =p\left(  a,b_{\perp}\right)  =\frac{1}{2}%
\sin^{2}\left(  a,b\right)  \tag{2}%
\end{equation}
and
\begin{equation}
p\left(  a_{\perp},b_{\perp}\right)  =p\left(  a,b\right)  =\frac{1}{2}%
\cos^{2}\left(  a,b\right)  , \tag{3}%
\end{equation}
where $p\left(  a_{\perp},b\right)  $ is the probability of detecting $\nu
_{1}$ in a polarization state perpendicular to $\mathbf{a}$ and $\nu_{2}$ in a
polarization state parallel to $\mathbf{b}$, and so on\textbf{. }Please, note,
for instance, that we can write\textrm{ }$p\left(  a_{\perp},b\right)
=p(a_{\perp})p(b\mid a_{\perp})$, where\textrm{ }$p(a_{\perp})=1/2$\textrm{
}is the probability of\textrm{ }$\nu_{1}$\textrm{ }being detected\textrm{
}at\textrm{ }$D_{1^{\prime}}$, and $p(b\mid a_{\perp})=\sin^{2}(a,b)$\textrm{
}is the probability of\textrm{ }$\nu_{2}$\textrm{ }being detected at\textrm{
}$D_{2}$\textrm{ }when\textrm{ }$\nu_{1}$\textrm{ }has been detected
at\textrm{ }$D_{1^{\prime}}$, forcing\textrm{ }$\nu_{2}$\textrm{ }into a state
perpendicular to $\mathbf{a}$. Similarly, assuming that NR-detections are
capable of inducing the reduction of the state vector, we can write, for
instance,\textrm{ }$p\left(  a,b_{\perp}\right)  =p(a)p(b_{\perp}\mid a)$,
where\textrm{ }$p(a)=1/2$\textrm{ }is the probability of\textrm{ }$\nu_{1}$
not being detected at $D_{1^{\prime}}$, and\textrm{ }$p(b_{\perp}\mid
a)=\sin^{2}(a,b)$\textrm{ }is the probability of\textrm{ }$\nu_{2}$\textrm{
}being detected at\textrm{ }$D_{2^{\prime}}$\textrm{ }when $\nu_{1}$\textrm{
}has not been detected at\textrm{ }$D_{1^{\prime}}$, but has not been detected
at\textrm{ }$D_{1}$\textrm{ }yet, even so forcing\textrm{ }$\nu_{2}$ into a
state parallel to\textrm{ }$\mathbf{a}$\textrm{. }On the other hand, if NR
detections are incapable of inducing the reduction of the state vector, we
must have
\begin{equation}
p^{\prime}\left(  a_{\perp},b\right)  =p\left(  a_{\perp},b\right)  =\frac
{1}{2}\sin^{2}\left(  a,b\right)  , \tag{4}%
\end{equation}%
\begin{equation}
p^{\prime}\left(  a_{\perp},b_{\perp}\right)  =p\left(  a_{\perp},b_{\perp
}\right)  =\frac{1}{2}\cos^{2}\left(  a,b\right)  , \tag{5}%
\end{equation}%
\begin{equation}
p^{\prime}\left(  a,b\right)  =\frac{1}{2}\alpha\left(  a,b\right)  \tag{6}%
\end{equation}
and
\begin{equation}
p^{\prime}\left(  a,b_{\perp}\right)  =\frac{1}{2}\left[  1-\alpha\left(
a,b\right)  \right]  , \tag{7}%
\end{equation}
where, in general, $\alpha\left(  a,b\right)  =p^{\prime}(b\mid a)\neq p(b\mid
a)=\cos^{2}\left(  a,b\right)  $, since, when $\nu_{1}$ is not detected at
$D_{1^{\prime}}$, $\nu_{2}$ (which, in this case, is detected before\textrm{
}$\nu_{1}$)\textrm{ }is not forced into a polarization state parallel to
$\mathbf{a}$ \textrm{[17]}.\ Naturally, when\textrm{ }$\nu_{1}$\textrm{ }is
detected at\textrm{ }$D_{1^{\prime}}$,\textrm{ }$\nu_{2}$\textrm{ }is forced
into a polarization state perpendicular to $\mathbf{a}$, and the usual quantum
mechanical results are obtained (eq. $(4)$\textrm{ }and\textrm{ }%
$(5)$)\textrm{. }It is interesting to observe that
\begin{equation}
p^{\prime}\left(  b\right)  =p^{\prime}\left(  a,b\right)  +p^{\prime}\left(
a_{\perp},b\right)  =\frac{1}{2}\left[  \alpha\left(  a,b\right)  +\sin
^{2}\left(  a,b\right)  \right]  \tag{8}%
\end{equation}
and
\begin{equation}
p^{\prime}\left(  b_{\perp}\right)  =p^{\prime}\left(  a,b_{\perp}\right)
+p^{\prime}\left(  a_{\perp},b_{\perp}\right)  =\frac{1}{2}\left[
1-\alpha\left(  a,b\right)  +\cos^{2}\left(  a,b\right)  \right]  . \tag{9}%
\end{equation}
Therefore, in this hybrid situation, in which\textrm{ }$\nu_{2}$ is forced
into a well-defined polarization state whenever $\nu_{1}$\textrm{ }is detected
at\textrm{ }$D_{1^{\prime}}$, and no forcing occurs if\textrm{ }$\nu_{1}%
$\textrm{ }is not detected at\textrm{ }$D_{1^{\prime}}$, we may have
$p^{\prime}\left(  b\right)  \neq p^{\prime}\left(  b_{\perp}\right)
$.\textrm{ }

To determine the disagreement between the NR-detection reduction of the state
vector (NR-detection collapse, for short)\ and the no-NR-detection
collapse\ approaches, I will introduce a local model \textrm{[6]}. This
artifice needs a brief explanation. I am assuming that whenever\textrm{\ }%
$\nu_{1}$\textrm{ }is\ not detected at\textrm{\ }$D_{1^{\prime}}$, but has not
yet been detected at\textrm{\ }$D_{1}$, nothing happens to\textrm{\ }$\nu_{2}%
$. In this case, supposing that\textrm{ }$\nu_{2}$\textrm{ }remains
unpolarized, we simply must have\textrm{ }$\alpha(a,b)=1/2$\textrm{. }However,
for the sake of completeness, I will also contemplate the possibility of
having a local hidden variable (LHV) connection between the entangled photons,
that is:\textrm{ }$\alpha(a,b)\equiv\alpha(a,b;\lambda)$, where\textrm{
}$\lambda$ stands for all possible LHV. In other words, there might be some
properties previously shared by the particles. In order to have space-like
events, we can remove the detours in the experiment represented in Fig.1;
however, the results would not change if we considered time-like events
instead, since the correlations are supposed to be determined only by the
previously shared properties. Using $(7)$ and $(6)$, the probabilities will be
given by
\begin{equation}
p_{L}\left(  a_{\perp},b\right)  =p_{L}\left(  a,b_{\perp}\right)  =p^{\prime
}(a,b_{\perp})=\frac{1}{2}\left[  1-\alpha\left(  a,b\right)  \right]
\tag{10}%
\end{equation}
and
\begin{equation}
p_{L}\left(  a_{\perp},b_{\perp}\right)  =p_{L}\left(  a,b\right)  =p^{\prime
}(a,b)=\frac{1}{2}\alpha\left(  a,b\right)  . \tag{11}%
\end{equation}
Hence, the correlation function will be
\[
E_{L}\left(  a,b\right)  =\frac{p_{L}\left(  a,b\right)  -p_{L}\left(
a,b_{\perp}\right)  -p_{L}\left(  a_{\perp},b\right)  +p_{L}\left(  a_{\perp
},b_{\perp}\right)  }{p_{L}\left(  a,b\right)  +p_{L}\left(  a,b_{\perp
}\right)  +p_{L}\left(  a_{\perp},b\right)  +p_{L}\left(  a_{\perp},b_{\perp
}\right)  }%
\]%
\begin{equation}
=2\alpha\left(  a,b\right)  -1. \tag{12}%
\end{equation}
Using the CHSHB inequality \textrm{[4, 16]}
\begin{equation}
\left\vert E_{L}\left(  a,b\right)  -E_{L}\left(  a,b^{\prime}\right)
+E_{L}\left(  a^{\prime},b\right)  +E_{L}\left(  a^{\prime},b^{\prime}\right)
\right\vert \leq2, \tag{13}%
\end{equation}
choosing $angle(a,b)=angle(a^{\prime},b)=angle(a^{\prime},b^{\prime})=\frac
{1}{3}angle(a,b^{\prime})=22.5^{\circ}$, and taking into account that
$E_{L}\left(  67.5^{\circ}\right)  =-E_{L}\left(  22.5^{\circ}\right)  $, we
obtain
\begin{equation}
4E_{L}\left(  22.5^{\circ}\right)  \leq2, \tag{14}%
\end{equation}
which, using $(12)$, leads to
\begin{equation}
\alpha\left(  22.5^{\circ}\right)  \leq\frac{3}{4}. \tag{15}%
\end{equation}
Using $(8)$ and $(9)$ we obtain
\begin{equation}
p^{\prime}\left(  b_{\perp}\right)  -p^{\prime}\left(  b\right)  =\frac{1}%
{2}-\alpha+\frac{1}{2}\cos45^{\circ}, \tag{16}%
\end{equation}
which, using $(15)$, leads to
\begin{equation}
p^{\prime}\left(  b_{\perp}\right)  -p^{\prime}\left(  b\right)  \geq
\frac{\sqrt{2}-1}{4}\approx0.1, \tag{17}%
\end{equation}
in disagreement with the quantum mechanical prediction $p\left(  b_{\perp
}\right)  -p\left(  b\right)  =0$. Naturally, if there is no classical
correlation, $\alpha=const.=1/2$, $p^{\prime}\left(  a,b\right)  =p^{\prime
}\left(  a,b_{\perp}\right)  =const.=1/4$, and instead of $(17)$ we will have
$p^{\prime}\left(  b_{\perp}\right)  -p^{\prime}\left(  b\right)  \approx
0.35$. But, in this case, the greatest disagreement is obtained choosing
$\mathbf{a=b}$, which, using $(9)$ and $(8)$, leads to $p^{\prime}\left(
b_{\perp}\right)  =3/4$ and $p^{\prime}\left(  b\right)  =1/4$, and, using
$(7)$, to $p^{\prime}(b,b_{\perp})=1/4\neq p(b,b_{\perp})=0$.

\bigskip

\textbf{III. The Real Situation}

\smallskip\medskip

In a real situation (assuming, for purposes of simplification, that all
detectors have the same efficiency, and that the two polarizers are
identical), instead of $(2)$ and $(3)$, we have \textrm{[16]}%

\begin{equation}
p(a,b)=p(a_{\bot},b_{\bot})=\frac{1}{4}\eta^{2}fg\left[  T_{+}^{2}+FT_{-}%
^{2}\cos2\left(  a,b\right)  \right]  \tag{18}%
\end{equation}
and%
\begin{equation}
p(a,b_{\bot})=p(a_{\bot},b)=\frac{1}{4}\eta^{2}fg\left[  T_{+}^{2}-FT_{-}%
^{2}\cos2\left(  a,b\right)  \right]  , \tag{19}%
\end{equation}
where $\eta$ is the detectors' efficiency; $f$ is the probability of the first
photon being collected; $g$ is the probability of the second photon being
collected when the first has been collected; $T_{\pm}=T_{\Vert}\pm T_{\bot}$,
where $T_{\parallel}(T_{\perp})$ is the transmission coefficient for light
polarized parallel (perpendicular) to the polarizer's orientation; and $F$
indicates the amount of correlation between the photons. Actually, unlike what
has been done in section II, the best procedure (from a practical point of
view) is to consider only the coincident detections in which $D_{1}$
clicks\textrm{,} since when $D_{1^{\prime}}$ clicks no NR detection
occurs.\textrm{ }From $(18)$ and $(19)$ we see that%
\begin{equation}
p_{c}(a)=p(a,b)+p(a,b_{\perp})=\frac{1}{2}\eta^{2}fgT_{+}^{2}, \tag{20}%
\end{equation}
where the subscript $c$ indicates that we are only considering situations in
which $\nu_{1}$ and $\nu_{2}$ are both detected (coincident detections).
Therefore,%
\begin{equation}
p(b\mid a)=\frac{p(a,b)}{p_{c}(a)}=\frac{1}{2}\left[  1+F\frac{T_{-}^{2}%
}{T_{+}^{2}}\cos2(a,b)\right]  \tag{21}%
\end{equation}
and%
\begin{equation}
p(b_{\perp}\mid a)=\frac{p(a,b_{\perp})}{p_{c}(a)}=\frac{1}{2}\left[
1-F\frac{T_{-}^{2}}{T_{+}^{2}}\cos2(a,b)\right]  , \tag{22}%
\end{equation}
where $p(b\mid a)\left[  p(b_{\perp}\mid a)\right]  $ is the probability of
$\nu_{2}$ being detected at $D_{2}\left[  D_{2^{\prime}}\right]  $ when
$\nu_{1}$ is detected at $D_{1}$, and $(18)$, $(19)$, and $(20)$ have been
used. Using $(21)$ and $(22)$ we obtain%
\begin{equation}
p(b\mid a)-p(b_{\perp}\mid a)=F\frac{T_{-}^{2}}{T_{+}^{2}}\cos2(a,b), \tag{23}%
\end{equation}
which can be written, using only directly observable quantities, as
\begin{equation}
\frac{N(a,b)-N(a,b_{\perp})}{N(a,b)+N(a,b_{\perp})}=F\frac{T_{-}^{2}}%
{T_{+}^{2}}\cos2(a,b), \tag{24}%
\end{equation}
where $N(a,b)[N(a,b_{\perp})]$ is the number of coincident detections at
$D_{1}$ and $D_{2}[D_{2^{\prime}}]$. Hence, using the data from the experiment
by Aspect, Grangier, and Roger \textrm{[16]} ($T_{\parallel}\approx0.950$,
$T_{\perp}\approx0.007$, and $F\approx0.984$), we see that%
\begin{equation}
\frac{N(a,b)-N(a,b_{\perp})}{N(a,b)+N(a,b_{\perp})}\approx0.696,\text{
\ \ }[angle(a,b)=22.5^{\circ}]. \tag{25}%
\end{equation}
On the other hand, in a real situation, if no NR-detection collapse occurs,
instead of $(6)$ and $(7)$ we must have%
\begin{equation}
p^{\prime}(a,b)=\frac{1}{4}\eta^{2}fg\left[  T_{+}^{2}+\beta\left(
a,b\right)  \right]  \tag{26}%
\end{equation}
and%
\begin{equation}
p^{\prime}(a,b_{\perp})=\frac{1}{4}\eta^{2}fg\left[  T_{+}^{2}-\beta\left(
a,b\right)  \right]  , \tag{27}%
\end{equation}
where, in general, $\beta\left(  a,b\right)  \neq FT_{-}^{2}\cos2(a,b)$.
Introducing a local model satisfying the conditions%
\begin{equation}
p_{L}(a,b)=p_{L}(a_{\perp},b_{\perp})=p^{\prime}(a,b) \tag{28}%
\end{equation}
and%
\begin{equation}
p_{L}(a,b_{\perp})=p_{L}(a_{\perp},b)=p^{\prime}(a,b_{\perp}), \tag{29}%
\end{equation}
instead of $(12)$ we obtain%
\begin{equation}
E_{L}(a,b)=\frac{\beta\left(  a,b\right)  }{T_{+}^{2}}, \tag{30}%
\end{equation}
which, using $(13)$, leads to%
\begin{equation}
\frac{\beta\left(  a,b\right)  }{T_{+}^{2}}\leq\frac{1}{2},\text{
\ \ }[angle(a,b)=22.5^{\circ}]. \tag{31}%
\end{equation}
From $(26)$ and $(27)$ we have%
\begin{equation}
p_{c}^{\prime}(a)=p^{\prime}(a,b)+p^{\prime}(a,b_{\perp})=\frac{1}{2}\eta
^{2}fgT_{+}^{2}. \tag{32}%
\end{equation}
Hence,%
\begin{equation}
p^{\prime}(b\mid a)=\frac{p^{\prime}(a,b)}{p_{c}^{\prime}(a)}=\frac{1}%
{2}\left[  1+\frac{\beta\left(  a,b\right)  }{T_{+}^{2}}\right]  \tag{33}%
\end{equation}
and%
\begin{equation}
p^{\prime}(b_{\perp}\mid a)=\frac{p^{\prime}(a,b_{\perp})}{p_{c}^{\prime}%
(a)}=\frac{1}{2}\left[  1-\frac{\beta\left(  a,b\right)  }{T_{+}^{2}}\right]
, \tag{34}%
\end{equation}
where $(26)$, $(27)$, and $(32)$ have been used. Therefore, using $(31)$, we
can write%
\begin{equation}
p^{\prime}(b\mid a)\leq\frac{1}{2}\left(  1+\frac{1}{2}\right)  \tag{35}%
\end{equation}
and%
\begin{equation}
p^{\prime}(b_{\perp}\mid a)\geq\frac{1}{2}\left(  1-\frac{1}{2}\right)  ,
\tag{36}%
\end{equation}
which leads to%
\begin{equation}
p^{\prime}(b\mid a)-p^{\prime}(b_{\perp}\mid a)\leq\frac{1}{2} \tag{37}%
\end{equation}
and to%
\begin{equation}
\frac{N^{\prime}(a,b)-N^{\prime}(a,b_{\perp})}{N^{\prime}(a,b)+N^{\prime
}(a,b_{\perp})}\leq0.5,\text{ \ \ }[angle(a,b)=22.5^{\circ}], \tag{38}%
\end{equation}
in strong disagreement with $(25)$. Assuming, in Fig.1, the distance
from\textrm{\ }$S$ to the first polarizer as being equal to 1 meter, we can
easily see that the lengths of the detours for\textrm{\ }$\nu_{2}$
and\textrm{\ }$\nu_{1}$,\ respectively, can be of approximately 2 and 4 m,
which can easily be accomplished using optical fibers.

\bigskip\textbf{IV. Discussion}

\bigskip

As emphasized at the beginning of this paper, NR-detection collapse\ is far
from being a trivial fact, and as such deserves to be experimentally
investigated. As already stressed in ref. 5 and 6, an interesting result is
that, if NR-detection collapse\ does not occur, then FTL communication becomes
possible, at least in principle, provided we are able to establish which
photon of an entangled pair is really\ detected first \textrm{[18]}.\textrm{
}For instance, by monitoring the number of detections on the right-hand side
of the experimental apparatus (Fig.1), it would be possible to know whether
detector $D_{1}$ on the left-hand side has been \textquotedblleft
removed\textquotedblright\ (that is, if a detour has been introduced) or not
\textrm{[19], }since it follows from\textrm{ }$(17)$\textrm{ }that\textrm{
}$p^{\prime}(b_{\perp})\neq p^{\prime}(b)$. This is in agreement with
Svetlichny's arguments\textrm{\ [20] }supporting the standpoint according to
which a causal theory (i.e. without superluminal communication) implies formal
state collapse. Although, at first sight, FTL communication seems inconsistent
with special relativity, things may not be so simple. As shown in the appendix
(and surprising as it may sound), no causal paradox necessarily arises from
superluminal signaling, even maintaining the Lorentz transformations.

The second possible result\ is that, as expected, NR-detection
collapse\ occurs. In this case, from an ontological standpoint (namely,
assuming that the collapse involves a change in the physical properties of the
system),\textrm{ }there seem to be\ two alternatives \textrm{[21]}: (a)
adopting a pilot wave interpretation a la de Broglie-Bohm, we can assume that
$\nu_{2}$\textrm{ }is forced into a well-defined polarization state when
$\nu_{1}$\textrm{ }is split in the polarizer into a \textquotedblleft
full\textquotedblright\ wave and an \textquotedblleft empty\textquotedblright%
\ wave \textrm{[}$\mathrm{22}$\textrm{]}; (b) ascribing an objective meaning
(or substance, so to speak)\ to the probability amplitude \textrm{[}%
$\mathrm{23}$\textrm{]}, we have to treat O\ and NR detections on the same
footing; that is, each time an O\ detection (or more generally, a photon
absorption)\ fails to occur in one branch of the experiment, the probability
amplitudes associated to the other branches are altered (as a consequence of
the changes experienced by the physical system), and automatically
adjusted:\ instead of photon absorption, we have the absorption (actually, the
redistribution) of a probability amplitude, which corresponds to an NR detection.

The above remarks raise the question about the possibility of quantum
mechanics being superseded. The most immediate thought that comes to mind is
of a new theory that keeps many essential features of the \textquotedblleft
old\textquotedblright\ one. However, Bohr's model of the atom, in which
accelerated charges do not necessarily radiate, is a good example of the fact
that it is not always wise to become too attached to the prevailing views,
even though these views will eventually play an important role in a more
elaborate formulation of the new theory \textrm{[}$\mathrm{24}$\textrm{]}. NR
detections (if capable of inducing a collapse of the wavefunction, as
expected)\textrm{ }strongly suggest that no amplification is involved in the
reduction of the state vector; in this case, adopting an ontological point of
view\textrm{ [12],}\ we may conjecture that at a\textrm{ }more fundamental
level than that ruled by quantum mechanics some as yet unknown processes take
place which are responsible for the so called \textquotedblleft actualization
of potentialities\textquotedblright.

\bigskip\textbf{V. Conclusion}

\bigskip

In this paper it has been assumed that the ordinary (O) detection of a photon
of a polarization entangled pair forces its twin into a well-defined
polarization state. On the other hand, the consequences of the conjecture that
this might not be true for null-result (NR) detections have been investigated.
More specifically, it has been pointed out that no noncontroversial collapse
inducing NR experiment has been performed so far, and a suggestion to remedy
this situation has been presented\textrm{. }Naturally,\ if the proposed
experiment was conducted and had a surprising result (i.e. if it was found
that NR-detection does not reduce the state vector), this would have
significance for ordinary quantum theory, and, furthermore, it would allow us
to infer an important thing about nature, viz. that, as stressed in the paper,
superluminal communication would be possible in principle. As shown in the
Appendix, this does not necessarily imply a causal paradox or the abandonment
of Lorentz transformations.\textrm{ }Actually, I would be very surprised if it
was found that NR-detections do not reduce the state vector. However, I also
think that we can only be certain about this point after a conclusive
experiment has been performed. As stressed in the previous section, whatever
the outcome of the experiment that is being proposed, it will have important
consequences for the foundations of QM.

\bigskip

\textbf{Appendix:}\textrm{ }\textbf{Superluminal Signaling Without Causal
Paradox}

\smallskip\medskip

The prevalent opinion is that the idea of faster-than-light (FTL)
communication leads to causal paradoxes. Although, in principle, the
introduction of a privileged frame would circumvent this problem, it is not
immediately obvious how to conciliate FTL\ communication with the Lorentz
transformations. Here it is shown--and, as far as I know, this is a new
result--how to do this by breaking the Lorentz symmetry; that is, although the
Lorentz transformations remain valid, the equivalence between passive and
active transformations is violated when superluminal communication is considered.

\smallskip Let us suppose that behind EPR correlations there is an FTL
interaction that can be used for superluminal communication, an idea
entertained by Bohm \textrm{[8]}. As we will see, this does not necessarily
lead to causal paradoxes, provided that we assume the existence of a preferred
frame (an aether, as imagined by Bell \textrm{[19]}) in which the speed of the
FTL interaction is a constant, that is, it is always the same, independently
of the motion of the \textquotedblleft source\textquotedblright\ (or, equally,
of the reference frame in which the experiment is being performed). Let us
consider a pair of reference frames, $\mathbf{S}$ and $\mathbf{S}^{\prime}$,
in the standard configuration, where $\mathbf{S}$ is the preferred\textrm{
}frame and $\mathbf{S}^{\prime}$ moves with velocity\textrm{ }$v<c$ along
the\textrm{ }$x$ axis. Assuming that the Lorentz transformations%
\begin{equation}
x^{\prime}=\gamma\left(  x-vt\right)  , \tag{a.1}%
\end{equation}%
\begin{equation}
t^{\prime}=\gamma\left(  t-\frac{v}{c^{2}}x\right)  , \tag{a.2}%
\end{equation}%
\begin{equation}
x=\gamma\left(  x^{\prime}+vt^{\prime}\right)  , \tag{a.3}%
\end{equation}
and%
\begin{equation}
t=\gamma\left(  t^{\prime}+\frac{v}{c^{2}}x^{\prime}\right)  , \tag{a.4}%
\end{equation}
connect the $\mathbf{S}$ and $\mathbf{S}^{\prime}$ coordinates (with\textrm{
}$\gamma=1/\sqrt{1-v^{2}/c^{2}}$), we derive%
\begin{equation}
u^{\prime}=\frac{u-v}{1-\frac{vu}{c^{2}}} \tag{a.5}%
\end{equation}
and%
\begin{equation}
u=\frac{u^{\prime}+v}{1+\frac{vu^{\prime}}{c^{2}}} \tag{a.6}%
\end{equation}
for the velocities \textrm{[}$\mathrm{25}$\textrm{]}.

Let us, initially, see how the causal paradox arises in special relativity (in
which there is no preferred frame and\textrm{ }$\mathbf{S}$ and $\mathbf{S}%
^{\prime}$ are\textrm{ }equivalent). Let the positive quantity\textrm{
}$\overline{u}>c$ represent the superluminal signal speed in $\mathbf{S}$.
From $(a.5)$, we see that if\textrm{ }$u=\overline{u}$, we can choose\textrm{
}$v$ so as to have\textrm{ }$v\overline{u}/c^{2}>1$, which leads to\textrm{
}$u^{\prime}<0$\textrm{ }(with\textrm{ }$\left\vert u^{\prime}\right\vert >c$
but\textrm{ }$\neq\overline{u}$). Therefore, in $\mathbf{S}^{\prime}$\textrm{
}the signal propagates backwards. Similarly, from\textrm{ }$(a.6)$ we see
that, if\textrm{ }$u^{\prime}=-\overline{u}$, we can choose a\textrm{ }$v$
that leads to\textrm{ }$u>0$\textrm{ }(with\textrm{ }$u>c$\textrm{
}and\textrm{ }$\neq\overline{u}$). That is, in $\mathbf{S}$ the direction of
propagation of the signal is reversed. It is this change of direction when we
go from $\mathbf{S}$ to\textrm{ }$\mathbf{S}^{\prime}$, and then from
$\mathbf{S}^{\prime}$\textrm{ }to $\mathbf{S}$, that is at the origin of the
causal paradox. To see this, let us consider a\textrm{ }superluminal signal
emitted from\textrm{ }$x_{0}=0$, at instant\textrm{ }$t_{0}=0$, and
reaching\textrm{ }$x_{1}>0$ at instant\textrm{ }$t_{1}$ given by%
\begin{equation}
t_{1}=\frac{x_{1}}{\overline{u}} \tag{a.7}%
\end{equation}
in $\mathbf{S}$. In $\mathbf{S}^{\prime}$, the signal is transmitted
from\textrm{ }$x_{0}^{\prime}=0$, at instant\textrm{ }$t_{0}^{\prime}=0$,
reaching\textrm{ }$x_{1}$ at instant%
\begin{equation}
t_{1}^{\prime}=\gamma\left(  t_{1}-\frac{v}{c^{2}}x_{1}\right)  =\gamma\left(
1-\frac{v\overline{u}}{c^{2}}\right)  \frac{x_{1}}{\overline{u}}, \tag{a.8}%
\end{equation}
according to\textrm{ }$(a.2)$ and $(a.7)$. We see that\textrm{ }$v\overline
{u}/c^{2}>1\rightarrow t_{1}^{\prime}<0$. Therefore, in $\mathbf{S}^{\prime}%
$\textrm{ }the signal reaches\textrm{ }$x_{1}$ before it\textrm{ }is sent
from\textrm{ }$x_{0}$\textrm{ }(actually, the signal is seen to propagate
from\textrm{ }$x_{1}$\textrm{ }to\textrm{ }$x_{0}$).\textrm{ }But this does
not yet represent a paradox, since no contradiction ($A$\textrm{ }and\textrm{
}$\lnot A$, for instance) is occurring. Let us then\textrm{ }determine the
point\textrm{ }$x_{1}^{\prime}$\textrm{ }in $\mathbf{S}^{\prime}$\textrm{
}that coincides with\textrm{ }$x_{1}$\textrm{ }at the instant at which the
signal arrives. Using\textrm{ }$(a.1)$ and\textrm{ }$(a.7)$, we obtain%
\begin{equation}
x_{1}^{\prime}=\gamma\left(  x_{1}-v\frac{x_{1}}{\overline{u}}\right)
=\gamma\left(  1-\frac{v}{\overline{u}}\right)  x_{1}. \tag{a.9}%
\end{equation}
An observer at\textrm{ }$x_{1}^{\prime}$\textrm{ }can then send a return
signal with\textrm{ }$u^{\prime}=-\overline{u}$\textrm{ }that will take the
time of%
\begin{equation}
\delta t^{\prime}=\frac{x_{1}^{\prime}}{\overline{u}}=\gamma\left(  1-\frac
{v}{\overline{u}}\right)  \frac{x_{1}}{\overline{u}} \tag{a.10}%
\end{equation}
to arrive at\textrm{ }$x_{0}^{\prime}$. This can lead to a paradox if%
\begin{equation}
t_{1}^{\prime}+\delta t^{\prime}<0, \tag{a.11}%
\end{equation}
that is, if the return signal reaches the origin of $\mathbf{S}^{\prime}%
$\textrm{ }before\textrm{ }$t_{0}^{\prime}$, namely before the first signal
has been sent. This enables an observer in this region, after receiving the
return signal, to inform another observer, at the origin of $\mathbf{S}$, not
to send the signal. As a consequence, if the signal is sent, it is possible to
send a return signal to impede the emission of the signal. That is, the signal
would be sent\textrm{ }and not sent at the same time! Let us see the
condition\textrm{ }$v$\textrm{ }would have to fulfil. From\textrm{ }$(a.11)$,
$(a.10)$,\textrm{ }and\textrm{ }$(a.8)$, we obtain%
\begin{equation}
\gamma\left(  1-\frac{v\overline{u}}{c^{2}}\right)  \frac{x_{1}}{\overline{u}%
}+\gamma\left(  1-\frac{v}{\overline{u}}\right)  \frac{x_{1}}{\overline{u}}<0,
\tag{a.12}%
\end{equation}
which leads to%
\begin{equation}
v>\frac{2\overline{u}}{1+\frac{\overline{u}^{2}}{c^{2}}}. \tag{a.13}%
\end{equation}
Since the right-hand side of\textrm{ }$(a.13)$\textrm{ }is always smaller
than\textrm{ }$c$, it is always possible to find a\textrm{ }$v$\textrm{ }that
satisfies the above condition; therefore, we would indeed have a paradox.

Now let us see how the existence of a preferred frame in which the
superluminal speed is a constant does not lead to a causal paradox. Instead
of\textrm{ }$(a.10)$, we have%
\begin{equation}
\delta t^{\prime}=\frac{x_{1}^{\prime}}{-\overline{u}^{\prime}}=-\gamma\left(
1-\frac{v}{\overline{u}}\right)  \frac{x_{1}}{\overline{u}^{\prime}},
\tag{a.14}%
\end{equation}
where the velocity of the return signal (using $(a.5)$)\textrm{ }is%
\begin{equation}
\overline{u}^{\prime}=\frac{-\overline{u}-v}{1+\frac{v\overline{u}}{c^{2}}}.
\tag{a.15}%
\end{equation}
The condition to have a causal paradox is then%
\begin{equation}
\gamma\left(  1-\frac{v\overline{u}}{c^{2}}\right)  \frac{x_{1}}{\overline{u}%
}-\gamma\left(  1-\frac{v}{\overline{u}}\right)  \frac{x_{1}}{\overline
{u}^{\prime}}<0, \tag{a.16}%
\end{equation}
where\textrm{ }$(a.14)$, $(a.11)$, and\textrm{ }$(a.8)$\textrm{ }have been
used. From\textrm{ }$(a.15)$ and\textrm{ }$(a.16)$\textrm{ }we obtain%
\begin{equation}
v>c, \tag{a.17}%
\end{equation}
which contradicts our initial assumption that the velocity of reference frame
$\mathbf{S}^{\prime}$\textrm{ }is slower than the velocity of light. As a
consequence, there can be no causal paradox.

Please note that the superluminal interaction that is being considered here
breaks the Lorentz symmetry, since the active transformation that would
correspond to the passive transformation does not exist (we have a similar
situation in the case of the violation of parity, in which some
mirror-reflected phenomena have no counterpart in the real world); that is, to
describe the FTL experiment from a frame that moves with velocity\textrm{
}$-\mathbf{v}$ relative to the preferred frame (passive transformation) is not
the same as to stay in the preferred frame and describe an FTL\ experiment in
which the experimental apparatus moves with velocity\textrm{ }$\mathbf{v}%
$\textrm{ }(active transformation).

\bigskip

\noindent\textbf{Acknowledgment: }I thank Paulo Henrique Souto Ribeiro for
helpful conversations.


\begin{thebibliography}{99}
\bibitem {1}The terms \textquotedblleft null\textquotedblright\ measurement,
\textquotedblleft null-result\textquotedblright\ measurement, and
\textquotedblleft negative-result\textquotedblright\textrm{ }measurement are
also used.
\bibitem {10}It may sound strange to consider the lack of detection as an
event; however, taking into account the context, there seems to be no reason
for any confusion.
\bibitem {17}To avoid any conceptual confusion, it is important to distinguish
between the terms\textit{ collapsing }and\textrm{ }\textit{forcing }in the
present context. Although\textrm{ }$\nu_{2}$ is not forced into a well-defined
polarization state (and, therefore, no \textquotedblleft
ordinary\textquotedblright\ reduction or collapse of the state vector occurs),
$\nu_{1}$\textrm{ }can no longer be detected at\textrm{ }$D_{1^{\prime}}$, as
a consequence, equation (1) is not valid. More\ specifically,\ $\nu_{1}$ can
no longer be detected in state\ $\mid a_{\perp}\rangle$, otherwise, we would
have the strange\ situation in which,\ by introducing\ a second
polarizer\ after the first\ one with the same\ orientation,\textrm{\ }$\nu
_{1}$\textrm{ }could be transmitted at the first polarizer and reflected at
the second. Whether this should be considered a sort of collapse, I prefer, by
the moment, to leave it as an open question. It is also important to notice
that, when writing the probabilities\textrm{ }$p(b\mid a)$\textrm{
}and\textrm{ }$p^{\prime}(b\mid a)$, we are assuming that whenever\textrm{
}$\nu_{1}$\textrm{ }is not detected at\textrm{ }$D_{1^{\prime}}$\textrm{ }it
will necessarily be detected at $D_{1}$. Otherwise, we would have more
detections at $D_{1^{\prime}}$\textrm{ }than at\textrm{ }$D_{1}$\textrm{.}
\end{thebibliography}
\end{document}